\documentclass{ws-ijmpb}

\begin{document}

\markboth{S. Malinin and  T. Nattermann} {
Transport in disordered Luttinger liquids}

%
\catchline{}{}{}{}{}
%

\title{TRANSPORT IN  LUTTINGER
LIQUIDS WITH STRONG IMPURITIES  }

\author{SERGEY MALININ}

\address{Department of
Chemistry,  Wayne State University\\
5101 Cass Avenue, Detroit,
Michigan 48202, USA\\
malinin@chem.wayne.edu }

\author{THOMAS NATTERMANN}

\address{Institut f\"ur Theoretische Physik, Universit\"at zu
K\"oln\\
Z\"ulpicherstr. 77, D-50937 K\"oln, Germany\\
natter@thp.uni-koeln.de}

\maketitle

\begin{abstract}
The tunnel current of a Luttinger liquid
with a finite density of strong impurities
is calculated using an instanton approach.
For very low temperatures $T$ or electric
fields $E$ the (nonlinear) conductivity is
of variable range hopping (VRH) type as
for weak pinning. For higher temperatures
or fields the conductivity shows power law
behavior corresponding to a crossover from
multi- to single-impurity tunneling. For
even higher $T$ and not too strong pinning
there is a second crossover to weak
pinning. The determination of the position
of the various  crossover lines both for
strong and weak pinning allows the
construction of the global regime diagram.
\end{abstract}


\section{Introduction}
$1$D electron systems exhibit a number of peculiarities which
destroy the familiar Fermi-liquid behavior known from higher
dimensions. Main reason is the geometrical restriction of the motion
in $1$D where electrons cannot avoid each other. As a consequence
excitations are plasmons similar to sound waves in solids. The
corresponding phase is called a Luttinger liquid (LL)
 \cite{FiGl96,Gi03}.
Renewed interest in LLs arises from progress in manufacturing narrow
quantum wires with a few or a single conducting channel. Examples
are carbon nanotubes \cite{CuZe04}, polydiacetylen \cite{AlLe04},
quantum Hall edges \cite{KaSt00} and semiconductor cleave edge
quantum wires \cite{AuYa02}.

From a theoretical point of view 1D quantum wires allow the
investigation of the interplay of interaction and disorder effects
since short range interaction can be treated already within a
harmonic bosonic theory \cite{Ha81}. Central quantity is the
interaction parameter $K$ which plays  the role of a dimensionless
conductance of a clean LL \cite{KaFi92,FiGl96}. The effect of
disorder on transport in LLs has been so far considered in two
limiting cases:

(i) The effect of a \emph{{single
impurity}} was considered in
\cite{KaFi92,GRS92,FuNa93,YuGlMa94}. Here
the conductance depends crucially on $K$.
Impurities are irrelevant for attractive
($K>1$) and strongly relevant for
repulsive interaction ($K<1$),
respectively.
 For finite voltage $V$ and
 $K<1$,
  the conductance is $\sim
  V^{\frac{2}{K}-2}\,\,\,$
  \cite{KaFi92}.
These considerations can be extended to
two impurities. Depending on the applied
gate
 voltage, Coulomb blockade
 effects may give rise to
 resonant tunneling  \cite{KaFi92,FuNa93}.

(ii) In the opposite case of a
\emph{finite density of \emph{weak}
impurities}, (Gaussian) disorder is a
relevant perturbation for $K<{3}/{2}$
leading to the  localization of electrons.
For weak external electric field $E$ the
conductivity is highly nonlinear: $\sigma
(E)\sim e^{-c /\sqrt{E}}$
\cite{Sh72,NaGiDo03,MaNaRo04,FoKe05}. At
low but finite temperatures $T$ this
result goes over into the VRH expression
for the linear conductivity $\sigma \sim
e^{-c'/\sqrt{T}}$
\cite{NaGiDo03,MaNaRo04,FoKe05,Mo74,FoTeSh04}.
At higher temperatures there is a
crossover to $\sigma \sim T^{2-2K}$
\cite{GiSc88}.

\begin{figure}[htb]
\includegraphics[width=01.0\linewidth]
{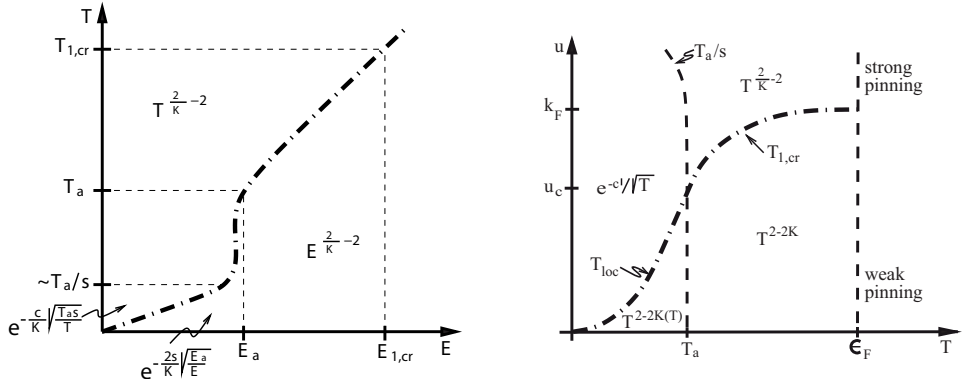} \caption{{Left}: Field
and temperature dependence of the
conductivity in the various regions of the
$T-E$ plane. $T_a, T_{1,\mathrm{cr}}, s,
E_a$ and $E_{1,\mathrm{cr}}$ are explained
in the text. The region $T_a>T>T_a/s$,
$E<E_a$ is characterized by activated
behavior $\sigma\sim
e^{-T_a/T}\sinh(\frac{Ea}{T})\frac{T}{Ea}$
Right: $u-T$ phase diagram of the linear
conductivity of disordered LLs. For strong
pinning, $u>u_c\sim k_F(k_Fa)^{K-1}$ and
$T<T_{1,\mathrm{cr}}\sim
T_a({u}/{u_c})^{{1}/({1-K})}$, $\,\,T_a/s$
separates the VRH from the single impurity
hopping regime.  For $T>T_{1,\mathrm{cr}}$
impurities become weak. For weak pinning,
$u<u_c$, $T_{\mathrm{loc}}\sim
T_a({u}/{u_c})^{{2}/({{3}-2K})} $
separates VRH from renormalized power law
behavior. For $T>T_a$ the power law is
unrenormalized. } \label{sigma}
\end{figure}

On the contrary,  much less is known in
the case of  a \emph{finite density of
\emph{strong} pinning centers}
\cite{GiMa96,Gi03} which  we will address
in
 the present paper.
In particular we determine both the
temperature and electric field dependence
of the (nonlinear) conductivity for this
case in a broad temperature and electric
field region. The main results of the
paper are the conductivities
(\ref{eq:large_E}), (\ref{eq:large_T}),
(\ref{eq:small_E}) and (\ref{eq:small_T})
as well as the crossover behavior
summarized in Fig. \ref{sigma}.

\section{Model and Instantons}
Starting point of our calculation is the
action of interacting electrons subject to
an external uniform electric field $E$ and
strong pinning centers. In bosonized form
the  action takes the form
\begin{equation}
  S =  \frac{\hbar}{2\pi K }
    \int\limits_0^{L} \int\limits_{0}^{\lambda_T}dx dy
    \Big\{(\partial_{y}\varphi)^{2}+
    (\partial_{x}\varphi + fx)^{2} -\sum\limits_{i=1}^N u \delta(x-x_i)
\cos(2\varphi+2k_Fx_i)\Big\}\,
\label{eq:n-zfr7}\end{equation} The phase
$\varphi(x)$ is related to the electron
density $\rho(x)=\pi^{-1}
(k_F+\partial_x\varphi)(1 + 2
\cos(2\varphi +2k_Fx))$. $k_F$ is the
Fermi wave vector,  $\tau={y}/{v}$ and
$f={FK}/{v\hbar} $. $v$ and
$\lambda_T=\hbar v/T$ denote the plasmon
velocity and the thermal de Broglie wave
length, respectively.

The phase field between the impurities can now be easily integrated
out leaving only its values $\varphi(x_i,y)\equiv\phi_i(y)$
\emph{at} the impurity sites $x_i$ which are assumed to be randomly
distributed. The action  can then be expressed in terms of Fourier
components  $\phi_i(y) ={\lambda_T}^{-1}\sum_{\omega_n}
\phi_{i,{\omega_n}}e^{-i\omega_n y}$, $\omega_n=2\pi n/\lambda_T$.
Thus
\begin{eqnarray}
\label{eq:fullaction} S=\frac{\hbar}{2\pi
K}\sum\limits_{i=0}^N\bigg\{
&&\hspace{-0.2cm}
\sum\limits_{\omega_n}\frac{\omega_n}{\lambda_T}
\left(\frac{|\phi_{i+1,\omega_n}-\phi_{i,\omega_n}|^2}
{\sinh \omega_n a_{i}} +
\left(|\phi_{i,\omega_n}|^2+
|\phi_{i+1,\omega_n}|^2\right)
\tanh\frac{\omega_n a_{i}}{2}
\right)\nonumber\\
&&-f(a_{i-1}+a_i)\phi_{i,0}
+u_{\mathrm{eff}} \int
dy\,\Big[1-\cos\big(2\phi_i(y)+2\pi\alpha_i\big)\Big]\bigg\}
\end{eqnarray}
where $a_i=x_{i+1}-x_i$ and $\alpha_i=k_Fx_i/\pi$. Since $k_Fa_i\gg
1$ below we will assume the $\alpha_i$ to be random phases but keep
the impurity distance $a\textbf{}_i$ approximately constant
$a_i\approx a $.

Next we consider the current  resulting from tunneling processes
between metastable states, assuming strong pinning and weak quantum
fluctuations, i.e. $K\ll 1$. The tunneling process starts from a
classical \emph{metastable} configuration $\tilde\phi_i$ which
minimizes the impurity potential for all values of $y$, $E=0$. Hence
$\tilde\phi_{i}{}= \pi(n_i-\alpha_i)$ where $n_i$ is integer. Among
the many metastable states there is one (modulo $\pi$)  zero field
{\emph{ground}} state $\tilde\phi_{i}^{0}$
 where
 $n_i=n_{i}^{0}=\sum_{j\le
i}[\Delta\alpha_{j-1}]_G$ \cite{NaGiDo03}.
Here
$\Delta\alpha_{j}=\alpha_{j+1}-\alpha_{j}$
and  $[\alpha]_G$ denotes the closest
integer
 to $\alpha$.
A new  metastable state follows from the ground state by adding
integers $q_i=\pm 1$ to the $n_i^{0}$.

Next we
consider an \emph{instanton} configuration
 which connects the original
state $\tilde\phi_i$ with the new state
$\tilde\phi_i+\pi$,
 $n_i$ depends in
general on $y$. To be specific, we assume
a double kink configuration for the
instanton at each impurity site:
$\phi_i(y)=\tilde\phi_{i}+\pi\,,$ for $
\quad |y- y_i|< D_i-d,\,\,\,$ and $ \,\,
\phi_i(y)=\tilde\phi_{i}\,,$ for $
|y-y_{i}|> D_i+d$, with a linear
interpolation between the two values at
the kink walls in the regions
$\big||y-y_i|- D_i\big|<d$. $y_i\pm D_i$
is the kink/anti-kink position, $d\sim
1/u$ is the approximate width of the kinks
and $2 D_i$ their distance. 
It is plausible that in the saddle point configuration all $y_i$
will  be the same, an approximation  we will use in the following.
With $z_i=\pi D_i/a$ the instanton action can then be rewritten as
\begin{equation}\nonumber
 S_{\mathrm{I}}\approx \frac{2\hbar}{K}{\sum_{i}}\left\{\frac{
 \Delta\tilde\phi_i}{\pi}
 (z_{i+1}-z_i) - fa^2 z_i+ s
  +  \ln \left[\frac{\cosh ((z_{i+1}-z_i)/2)} {\cosh
((z_{i+1}+z_i)/2)} \tanh \frac{
z_i}{2}\cosh{ z_i}{} \right]
\right\}\nonumber
\end{equation}
where the sum goes only over impurities
with $z_i>0$.  $s$ is a constant  that
includes the core action of a kink and an
anti-kink: $s=\ln(Cau)\gg 1$, where $\ln
C/K\gg 1$.

For a given initial metastable state $\{\tilde\phi_i\}$,
$S_{\mathrm{I}}$
 is a function of the variational parameters
$\{ D_i\,;\,\, i=1,...,N\}$.  The nucleation rate $\Gamma$ and hence
the current $I$ is given by $
    I\propto \Gamma \propto
    \prod\limits_{i=0}^N \int_0^{i\infty}
dD_k\exp(- S/\hbar).
$
 Here we employ
 an approximate treatment in
 which we assume
$D_i\equiv D=az/\pi$ for $k<i\le k+m$ and
$D_i=0$ elsewhere, i.e. tunneling is
assumed to occur simultaneously through
$m$ neighboring impurities. The instanton
is then a rectangular object with
extension $ma$ and $2D$ in $x$ and $y$
direction, respectively. The instanton
action can then be written as
\begin{equation}\nonumber\label{eq:m-instanton}
    S_{\mathrm{inst}}
    =\frac{2\hbar}{K}\left\{z\sigma_m(k)+ \ln(1+e^{-2z})
    +m\Big(s+\ln \tanh
    \frac{z}{2}-z\frac{E}{E_a}\Big)\right\}\,.
\end{equation}
Here we introduced the dimensionless field
strength $fa^2/\pi={E}/{E_a}$ where
$E_a=1/(\kappa a^2)$, $\kappa=K/\pi\hbar
v$ denotes the compressibility.
$\sigma_m(k)=
\big(\nu_k(1)+\nu_{k+m}(-1)\big)/2$ plays
the role of a surface tension of the
vertical boundaries of the instanton where
$\nu_k(q)=q^2-2q\big(\Delta\alpha_k-
[\Delta\alpha_k]_G\big)$. In the ground
state $\sigma_m(k)$ is equally distributed
in the interval $0\le \sigma_m(k)<2$
\cite{MaNaRo04}.
 The second and the
third contribution in (\ref{eq:m-instanton}) result from the
horizontal boundaries of the instanton and include their surface
tension $ s/a$ and their attractive interaction. The last term
describes the volume contribution resulting from the external field.

In addition, we have to include a small
dissipative term $S_{\mathrm{bath}}=
\frac{2\hbar}{K}\,m \eta \,\ln z$ ,
$\eta\ll 1$, in the action in order to
allow for energy dissipation
\cite{MaNaRo04}. However, we will omit
$\eta$-dependent terms in all results
where they give only small corrections
(apart from possible pre-exponential
factors which we do not consider).

A necessary condition for tunneling is
$\partial S_{\mathrm{inst}}/\partial z<0$
for $z\to \infty$, i.e.
$\sigma_m(k)<m{E_a}/{E}$. The tunneling
probability follows from the saddle point
value of the instanton action where $z$
fulfils the condition $
\sigma_m(k)-m\frac{E}{E_a}+\tanh z -1+
\frac{m\,\eta }{z}+\frac{m}{\sinh z}=0\,.$

\section{Results and Conclusions}
We discuss now several special cases: (i) For sufficiently large
fields $E \gg E_a$ the saddle point is $z_s\approx \frac{E_a}{E} \ll
1$ which gives a tunneling probability $ \Gamma \propto
    ({{E}/{E_a}})^{\frac{2m}{K}-1}
    e^{-2ms/K}$.
The exponent $-1$ results from the
integration around the saddle point.
Because of small $K$ and correspondingly
large kink core action, tunneling through
single impurities ($m=1$) is preferred and
hence the nonlinear conductivity is given
by
\begin{equation}\label{eq:large_E}
    \sigma(E)\sim \left(
     {E}/{E_a}
     \right)^{\frac{2}{K}-2}e^{-2s/K},
     \,\,\,\,\,\,\,\,\,
     E_a< E<
     E_{1,\mathrm{cr}}=E_ae^s,
\end{equation}
in agreement with previous results for
tunneling through a single weak link
\cite{KaFi92} if we identify $e^{-s/K}\sim
t$ with the hopping amplitude $t$ through
the link.   The upper field strength for
the validity of this result can be
estimated from $D_su\equiv z_s a u<1$
since the instantons loose then their
meaning. Using $u\to
u_{\mathrm{eff}}\approx
k_F(u/k_F)^{1/(1-K)}$ we find
$E_{1,\mathrm{cr}}\sim (k_F/e\kappa
a)(u/k_F)^{1/(1-K)}$, which can be also
read off directly from (\ref{eq:large_E})
as $E_{1,\mathrm{cr}}\sim E_ae^s$.
Classically ($K=0$), $E_{1,\mathrm{cr}}$
corresponds to the case when the field
energy $Ea$ the electron gains by moving
to the next impurity is smaller than the
pinning energy $u/\kappa$.

At finite temperatures there is a
crossover to a temperature $T\approx EaK$
dependent conductivity
\begin{equation}\label{eq:large_T}
    \sigma(T,E)\propto
    \left(\frac{T}{T_a}\right)^{\frac{2}{K}-2}
    e^{-2s/K}\sinh(\frac{Ea}{T})\frac{T}{Ea}\,,\qquad
    EKa< T< T_{1,\mathrm{cr}}=T_ae^s
\end{equation}
when the instanton extension $2D_s$
reaches $\lambda_T$, i.e. for
$E<E_aT/T_a$. For temperatures higher than
$T_{1,\mathrm{cr}}$ isolated impurities
are weak. Following the arguments of
\cite{GiMa96} one expects in this region
$\sigma\sim T^{2-2K}$.

(ii) In the opposite case of weak fields,
$E\ll E_a$, tunneling happens
simultaneously through many impurities and
the saddle point is  $z_s\gg 1$. In this
case we can estimate the typical surface
tension as $\sigma_k(m)\approx 1/m$ for a
chosen pair of sites $k$ and $k+m$,
respectively \cite{MaNaRo04}.
 For very large values
of $m$ we can treat $m$ as continuous and
the saddle point condition gives
$m_s\approx \sqrt{{E_a}/{E}}\gg 1$ and
$z_s\approx sE_a/(2{E})$. The tunneling
probability and hence the current is
proportional to
\begin{equation}\label{eq:small_E}
    I\sim \sigma (E) \sim
    e^{-\frac{2s}{K}
    \sqrt{E_a/{E}}},\,\, E< E_a.
\end{equation}
If we write
    the result in the VRH form
    \cite{Mo74}
    $I\sim e^{-2ma/\xi_{\mathrm{loc}}}$
we can identify the localization length
$\xi_{\mathrm{loc}}\approx aK/s$ of the
tunneling charges.  There is a crossover
to a temperature dependent conductivity if
$\lambda_T< 2D_s$, i.e. for $E<
sTE_a/T_a<E_a$ where
\begin{equation}\label{eq:small_T}
    \sigma(T) \sim e^{-\frac{2}{K}
    \sqrt{{{c}s
    T_a}/{T}})}\sinh(\frac{Ea}{T})\frac{T}{Ea}\,, \qquad
    EKa/{s}< T<{T_a}/{s}
\end{equation}
Results (\ref{eq:small_E}) and
(\ref{eq:small_T}) are in agreement with
those obtained for weak pinning
\cite{NaGiDo03,MaNaRo04}.
(iii) If $m$ is not too large (e.g. for
large $a$) we have to take into account
the discreteness of $m$. An instanton
solution exists only for $m>\sqrt{E_a/E}$.
Since $S_{\mathrm{inst}}(z(m),m)$
 has always a negative derivative
with respect to $m$ at $m\to
\sqrt{E_a/E}+0$, but for reasonably large
values of $s$ the interval of $m$ with
negative derivative is much shorter than
$1$ and hence the optimal hopping length
$m_s(E)$ is the smallest integer exceeding
$\sqrt{E_a/E}$, which we denote as
$\big[\sqrt{E_a/E}\,\big]_{G+}$. To be
more realistic we have  to take into
account the randomness of the impurity
distances $a_i$ such that decreasing the
field (or the temperature), the current
jumps by a factor $\sim
e^{-2a_m/\xi_{\mathrm{loc}}}$. Clearly,
for long wires these
jumps will  average out. %

Finally, we briefly compare  the present
case of Poissonian strong disorder,
$u_{\mathrm{eff}}a\gg 1$ with the Gaussian
weak disorder, $u_{\mathrm{eff}}a\ll 1$
considered in
\cite{FuNa93,NaGiDo03,MaNaRo04,GiSc88}. In
the latter case $u$ and $a$ are sent
simultaneously to zero but the quantity
$u^2/a\sim \xi_0^{-3}\ll k_F^3$ is assumed
to be finite, $\xi_0$ denotes the bare
correlation length. Fluctuations on scales
smaller than $\xi_0$ renormalize $\xi_0\to
\xi\sim k_F^{-1}(\xi_0k_F)^{3/(3-2K)}$. At
low $T$ the conductivity is of variable
range hopping type (\ref{eq:small_T}) up
to a temperature $T_{\mathrm{loc}}= \hbar
v/\xi=T_a(u/u_c)^{2/(3-2K)}$ where
$u_c\approx k_F(ak_F)^{K-1}$. For higher
$T$ there is a direct crossover to $\sigma
\sim T^{2-2K(T)}$ where $K$ is now
renormalized by disorder fluctuations
\cite{GiSc88,GiMa96}. This renormalization
disappears only at  much higher $T_a\sim
\hbar v/ a$. Both weak and strong pinning
theories should  roughly coincide for
$u\to u_c\approx k_F(ak_F)^{K-1}$ where
$T_a\approx T_{1,\mathrm{cr}}\approx
T_{\mathrm{loc}}$ which is indeed the case
since $\xi\approx a$. In the strong
pinning region $\xi$ continues as $\xi\sim
a/s$.

Experimentally, a linear variable range
hopping conductivity has been seen in
carbon-nanotubes \cite{CuZe04} and
polydiacetylen \cite{AlLe04}.

\section*{Acknowledgements}
 The authors thank A. Altland, T.
Giamarchi, B. Rosenow, S. Scheidl for
useful discussions. This work is supported
by the SFB 608 of DFG. S.M. acknowledges
financial support of  RFBR under Grant No.
03-02-16173.

\end{document}